\documentclass[a4paper,11pt]{article}
\pdfoutput=1

\usepackage[T1]{fontenc}
\usepackage[top=3cm,left=2.5cm,right=2.5cm,bottom=3cm]{geometry}
\usepackage{ragged2e}
\usepackage{lmodern}
\usepackage{multicol}
\usepackage{amsmath}
\usepackage{amsthm}
\usepackage{amsfonts}
\usepackage{thmtools}
\usepackage{mathtools}
\usepackage{amssymb}
\usepackage{bm}
\usepackage{enumitem}
\usepackage{graphicx}
\usepackage[dvipsnames]{xcolor}
\usepackage[bottom]{footmisc}
\usepackage{hyperref}
\usepackage{csquotes}
\usepackage{subcaption}

\definecolor{MATblue}{HTML}{0072BD}
\definecolor{MATpurple}{HTML}{7E2F8E}
\hypersetup{
	pdftitle={On the Limiting Distribution of Sieve VAR($\infty$) Estimators in Small Samples},    
	pdfauthor={Giovanni Ballarin},  
	colorlinks=true,
	linkcolor=MATpurple,
	citecolor=MATpurple,
	urlcolor=MATpurple
}
\newcommand{\Paragraph}[1]{\paragraph{\normalfont\textsc{#1}}}

\newcommand{\E}{\mathbb{E}}

\newcommand{\inv}{^{-1}}
\newcommand{\abs}[1]{\left\lvert#1\right\rvert}

\newcommand{\norm}[1]{\lVert#1\rVert}

\newcommand{\Real}{{\mathbb{R}}}
\newcommand{\Complex}{{\mathbb{C}}}

\newcommand{\refp}[1]{(\ref{#1})}

\usepackage[authoryear,round,comma]{natbib}
\begin{document}
	
\title{\normalfont On the Limiting Distribution of \linebreak Sieve VAR($\infty$) Estimators in Small Samples}
\author{Giovanni Ballarin%
	\footnote{%
		E-mail: \texttt{{giovanni.ballarin@gess.uni-mannheim.de}}.
		I am thankful to Carsten Trenkler for his helpful comments. I would also like to thank Jonas Krampe for proposing the main intuition behind this note, and for discussing its practical implications.
	}\\%
	University of Mannheim%
}
\date{\today}
	
\maketitle
	
\renewcommand{\abstractname}{\vspace{-\baselineskip}} 
\begin{abstract}
	\textbf{Abstract:}
	When a finite order vector autoregressive model is fitted to VAR($\infty$) data the asymptotic distribution of statistics obtained via smooth functions of least-squares estimates requires care.  \cite{lutkepohlEstimatingOrthogonalImpulse1991} provide a closed-form expression for the limiting distribution of (structural) impulse responses for sieve VAR models based on the Delta method. Yet, numerical simulations have shown that confidence intervals built in such way appear overly conservative. In this note I argue that these results stem naturally from the limit arguments used in \cite{lutkepohlEstimatingOrthogonalImpulse1991}, that they manifest when sieve inference is improperly applied, and that they can be "remedied" by either using bootstrap resampling or, simply, by using standard (non-sieve) asymptotics.
\end{abstract}
	
\pagebreak
	
\section{Introduction}
	
The framework of vector sieve autogressions is theoretically useful because in any practical application of vector autoregessive (VAR) models the assumption that the underlying process depends on a finite number of lags is easily debatable. For example, the commonly used New Keynesian DSGE models of modern macroeconomic research often reduce to a VARMA specification \citep{KilianLutkepohl2017}: even in the case of MA invertibility, the correct model therefore would be equivalent to a VAR($\infty$) process. Fitting a VAR($p$) model to a VAR($\infty$) data-generating process whenever $p < \infty$ is however inevitable given the finite amount of data and computational power available at any point in time. The sieve framework is specifically tailored to studying this situation, and correct inference is its end goal.

Somehow remarkably, it often appears that sieve asymptotic results yield poor performance. As \cite{inoueBootstrappingSmoothFunctions2002}, p. 318 write, referring to \cite{lutkepohlEstimatingOrthogonalImpulse1991}:
\begin{displayquote}
	The delta-method interval tends to be much wider on
	average at longer horizons than the bootstrap interval. Although we know that, as $T$
	approaches infinity, the interval endpoints of L{\"u}tkepohl and Poskitt’s delta-method
	interval and of the bootstrap percentile interval will coincide, for $h > k$ and fixed $T$
	the intervals can be quite different. We also note that, for $h > k$ and fixed $T$, the
	conventional asymptotic theory for bootstrapping finite-lag order models appears to
	provide a better approximation than the bootstrap asymptotic theory for VAR($\infty$)
	models.\footnote{
		Here $h$ indicates the impulse response horizon, while $k$ is the VAR lag order. 
	} 
\end{displayquote}
The issue of interval length is significant here, because in practice a researcher would want to construct confidence (or error) intervals which are not systematically far too wide for a given nominal level. This can be especially important when setting up tests for statistics of interest.

\begin{figure}
	\centering
	\includegraphics[width=0.8\linewidth]{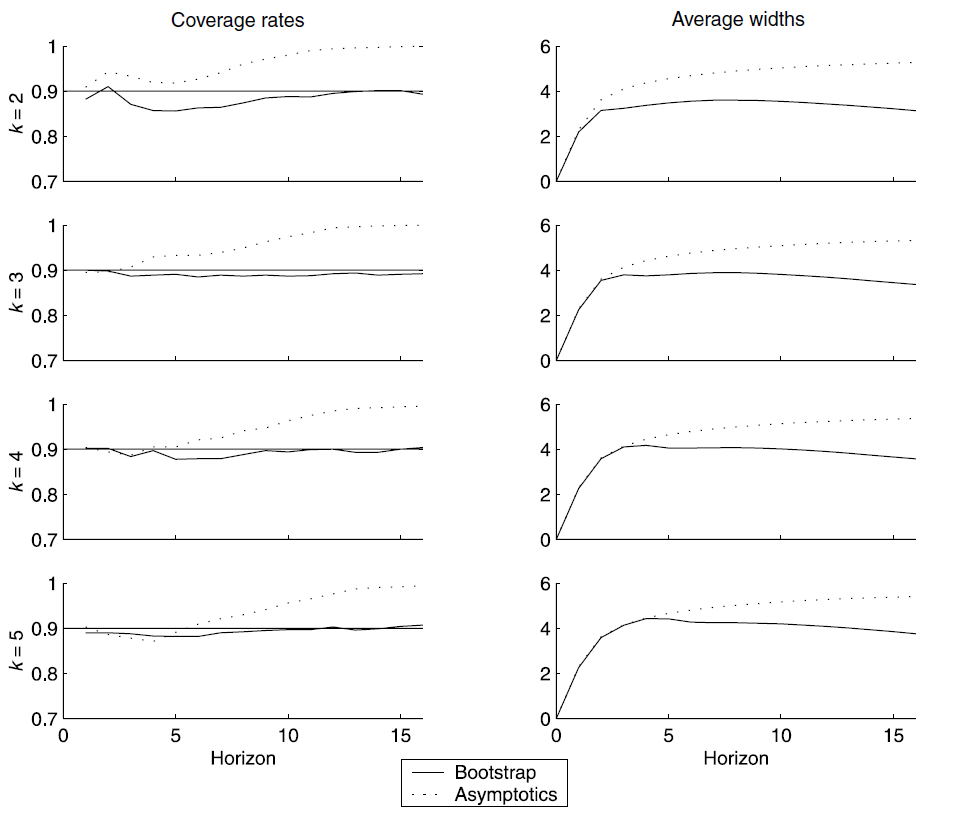}
	\caption{Monte Carlo simulation from \cite{inoueBootstrappingSmoothFunctions2002}, comparing coverage rates of sieve asymptotic (dashed) versus boostrap (solid) confidence intervals, nominal 90\% rate. $T = 200$ sample size, 1000 Monte Carlo replications.}
	\label{fig:inouekilian2002fig1}
\end{figure}

The main takeaway from \cite{inoueBootstrappingSmoothFunctions2002} can be seen in Figure \ref{fig:inouekilian2002fig1}, and refers to their Monte Carlo simulation exercise on impulse response function (IRF) inference.\footnote{
	Figure \ref{fig:inouekilian2002fig1} is reproduced from Figure 1, \cite{inoueBootstrappingSmoothFunctions2002}.
} As one may easily notice, it looks as if asymptotic sieve confidence intervals (CIs) are systematically over-conservative at long horizons, while bootstrap CIs clearly show more appropriate coverage properties. There are two issues with these results and this note tackles both of them in order to give a more correct understanding of the sieve method. This, in turn, should lead to a better practical use of sieve autoregressions.

Firstly, from a theoretical point of view, great care must be taken in comparing inference of sieve and non-sieve (including bootstrap) methods whenever the choice of $p$ is "free". The theory of sieve autoregressions \citep{lutkepohlAsymptoticDistributionMoving1988, lutkepohlEstimatingOrthogonalImpulse1991} makes specific assumptions on the relationship between $p$ and the sample size $T$. This relationship can not be ignored, as the sieve asymptotic theory hinges on them. This leads, in practice, to a meaningful differences in terms of finite sample properties of sieve estimators vis a vis finite-order VAR($p$) inference.

Secondly, when one is interested in impulse response functions, the sieve approach should only be used to study IRF horizons \textit{of at most $p$}. This tight link between $p$ and the maximal horizon for inference is built into the asymptotic theory itself. The same is true in general for any statistic that depends on any model parameter of the underlying VAR($\infty$) process that is \textit{not included} in the VAR($p$) estimation. Therefore, the results implied by Figure \ref{fig:inouekilian2002fig1} are misleading, because the "over-coverage" of sieve confidence intervals is a purely numerical artifact.

\Paragraph{Notation.}
 
The matrix norm $\norm{A}$ for $A \in \Real^{K\times K}$ is intended as the spectral norm, while $\norm{A}_F$ is the Frobenius norm. By the equivalence of matrix norms (for finite dimension $K$), i.e. $\norm{A} \leq \norm{A}_F \leq \sqrt{K} \norm{A}$, statements not depending on $K$ can feature either norm without loss of generality.

\section{The Autoregressive Sieve}

\Paragraph{Assumptions.}

Suppose $y_t$ is a causal, stationary, nondeterministic process of fixed dimension $K$ with stable VAR representation
\begin{equation*}
	y_t = \sum_{i=1}^\infty A_i y_{t-i} + u_t
\end{equation*}
and VMA representation 
\begin{equation*}
	y_t = \sum_{i=0}^\infty \Phi_i u_{t-i}, \qquad \Phi_0 = I_K
\end{equation*}
where the error term $u_t$ is i.i.d. distributed with $\E[u_t] = 0$, $\text{Var}[u_t] = \Sigma_u$ and $\E\abs{u_{it} u_{jt} u_{mt} u_{nt}} < \infty$ for $1 \leq i,j,m,n \leq K$. It also holds that 
\begin{equation*}
	\sum\limits_{i=1}^\infty \norm{A_i} < \infty, \qquad \sum\limits_{i=0}^\infty \norm{\Phi_i} < \infty
\end{equation*}
and 
\begin{equation*}
	\text{det}\left( \sum\limits_{i=0}^\infty \Phi_i z^i \right) \not= 0, \qquad \text{for} \abs{z} \leq 1, \ z\in \Complex
\end{equation*}

Let $p < \infty$ and consider the truncated VAR($p$) model given by coefficients $A_1, \ldots, A_p$. This model can be estimated from the data using least-squares, see \cite{lutkepohlNewIntroductionMultiple2005}. Define $J_p = [I_K, 0, \ldots, 0]$ as a $(K \times Kp)$ matrix, 
\begin{align*}
	\mathbb{A}_p & = \begin{bmatrix}
		A_1 & A_2 & \cdots & A_{p-1} & A_p \\
		I_K & 0 & \cdots & 0 & 0 \\
		0 & I_K & \cdots & 0 & 0 \\
		\vdots & \vdots & \ddots & \vdots & \vdots \\
		0 & 0 & \cdots & I_K & 0
	\end{bmatrix},
\end{align*}
and
\begin{align*}
	\Gamma_p & = \begin{bmatrix}
		\Gamma(0) & \Gamma(-1) & \cdots & \Gamma(-p+1) \\
		\Gamma(1) & \Gamma(0) & \cdots & \Gamma(-p+2) \\
		\vdots & \vdots & \ddots & \vdots \\
		\Gamma(p-1) & \Gamma(p-2) & \cdots & \Gamma(0) 
	\end{bmatrix}
\end{align*}
One can easily generalize the above display to a true VAR($\infty$) model by defining the infinite-dimensional matrices $J_\infty = [I_K, 0, \dots]$, $\mathbb{A}_\infty = \lim_{p \to \infty} \mathbb{A}_p$ and $\Gamma_\infty = \lim_{p \to \infty} \Gamma_p$.

It follows from this notation that $\Phi_i = J_p (\mathbb{A}_p)^i J_p'$ for all $p \geq i$. Equivalently, the MA coefficient matrices satisfy the recursive equation
\begin{equation}\label{eq:MA_coef}
	\Phi_i = \sum_{m=0}^{i-1} \Phi_m \, A_{i-m}
\end{equation}
where $\Phi_0 = I_K$.	

\subsection{Sieve Delta-method Asymptotics}

To make sieve asymptotic theory valid, \cite{lutkepohlEstimatingOrthogonalImpulse1991} and \cite{lutkepohlAsymptoticDistributionMoving1988} make the following assumption, which originates from \cite{lewisPredictionMultivariateTime1985}:

\paragraph{Assumption 1.}\label{A.1} The order $p$ of the fitted VAR model is such that $p \to \infty$, while respecting
\begin{itemize}
	\item[(i)] $p^3 / T \to 0$
	\item[(ii)] $\displaystyle \sqrt{T} \sum\limits_{i=p+1}^\infty \norm{A_i}_F \to 0$
\end{itemize}
as $T \to \infty$.

\paragraph{} Assumption 1 gives two bounds on the speed at which $p$ is allowed to grow with $T$. Condition (i) is an upper bound, while condition (ii) is a lower bound. It turns out that it is these conditions which closely relate to the "poor" finite sample properties of delta-method sieve inference results. In theory, (ii) can be the most problematic of the two. To see this, consider the simple case where $\norm{A_i}$ declines geometrically for some $i \gg 1$, that is $\norm{A_i} \asymp \alpha^i$ for $\alpha \in (0,1)$. Then
\begin{equation}\label{eq:1}
	\sqrt{T} \sum\limits_{i=p+1}^\infty \norm{A_i}  \asymp \sqrt{T} \left[ C_\alpha \, \alpha^{p} \right]
\end{equation}
because from the properties of geometric series it is immediate that
\begin{equation*}
	\sum_{i=0}^{\infty} c\, \alpha^i - \sum_{i=0}^{n} c\, \alpha^i = \alpha^{n+1} \left(\frac{c}{1 - \alpha}\right)
\end{equation*}
Therefore \refp{eq:1} follows by using appropriate constants, and assuming $p$ sufficiently large. Notice then that $p \propto \log(T)$ is indeed enough to satisfy $\sqrt{T} \left[ C_\alpha \, \alpha^{p} \right] \to 0$ as $T \to \infty$.\footnote{
	For example, consider $p = c_T\, \log(T)$ for $c_T > -(2 \log(\alpha))\inv$.
}

The core asymptotic result in \cite{lutkepohlAsymptoticDistributionMoving1988}, Theorem 1 (also used in \cite{lutkepohlEstimatingOrthogonalImpulse1991}), crucially hinges on the limit (p. 84)
\begin{equation}\label{eq:2}
	J_p (\mathbb{A}_p')^i \Gamma_p\inv (\mathbb{A}_p)^j J_p \ \to \ J_\infty (\mathbb{A}_\infty')^i \Gamma_\infty\inv (\mathbb{A}_\infty)^j J_\infty 
\end{equation}
as $p \to \infty$ to simplify the asymptotic variance of VMA sieve coefficient estimates. One may now ask, at what rate $T$ should grow \textit{as a function of} $p$ for Assumption 1 to hold? This is equivalent to asking the order of convergence of matrices $J_p$, $(\mathbb{A}_p')^i$ and $\Gamma_p\inv$ above to their limit \textit{strictly in terms of dimension}. 

For a linear growth of $p$ in \refp{eq:2}, if $p = o(T^{1/3})$ then clearly the sample size needs to increase at least as quickly as $T \propto p^3$. Yet even more unluckily, if it were that $p \propto \log(T)$ then $T \propto \exp(p)$. A simple example shows that these rates are rather extreme. In a model, to justify a shift from $p = 9$ to $p = 10$ would require, according to the above rates, that sample sizes increase by roughly $37\%$ and $170\%$, respectively. Since the rates are non-linear, the growth of $T$ with respect to $p$ becomes even harsher the more lags are considered.

\Paragraph{Inference horizon.} Closely related to the relationship between $p$ and $T$ is the choice of the maximal inference horizon $H$. It is common that the researcher is simply interested in the econometric analysis of $\{ \Phi_0, \ldots, \Phi_i, \ldots, \Phi_H \}$. Importantly then, the asymptotic distribution derived in \cite{lutkepohlAsymptoticDistributionMoving1988} is proven to be valid \textit{only for $i \leq p$}. In a given sample, the order $p$ therefore explicitly sets the maximal horizon at which one should be making inference on $\Phi_i$ using the sieve asymptotics, cf. \cite{lutkepohlAsymptoticDistributionMoving1988}, p. 83.

As \cite{lutkepohlEstimatingOrthogonalImpulse1991} remark,
\begin{displayquote}
	Finally, it is, perhaps, worth pointing out that since $\hat{\Phi}_{i,p}$ are generated via the difference equation [(\ref{eq:MA_coef})] they are functionally dependent [...] This means that the consideration of impulse responses or dynamic multipliers for lags greater than $p$ can provide no new independent information.\footnote{
		Notation has been adjusted from the original, where $\hat{\Phi}_{i,h} \equiv \hat{\Phi}_{i,p}$ and $h \equiv p$.
	}
\end{displayquote}
Constructing estimates of ${\Phi}_{p+1}, {\Phi}_{p+2}, \ldots$ via recursion, while possible, does not yield any information on the missing $A_{p+1}, A_{p+2}, \ldots$, and therefore sieve inference about $\hat{\Phi}_{p+1}, \hat{\Phi}_{p+2}, \ldots$ is not justified by theory. This is because by fitting a VAR($p$) model for $p < \infty$, the estimate $\hat{\Phi}_{i}$ for $i > p$ contains a (non-estimable) error term due to ignoring $A_i$ and this error does not vanish when \textit{only $T$} grows. It is true that for any \textit{fixed} $i \in \mathbb{N}$, for $T$ large enough Assumption \ref{A.1} ensures that eventually $A_i$ will be included in the estimation, so that such error becomes negligible in the limit $p \to \infty$. The problem is that, in any given data sample, $p$ will be chosen finite, and therefore, to be credible, any sieve inference should be strictly limited to statistics that depend \textit{only on the information until lag $p$}. This means one should avoid making sieve inference on impulse responses whenever $H \geq i > p$.

The alternative for the researcher is to believe that the VAR($p$) model captures all the relevant information regarding $\Phi_i$ even when $i > p$. The correct approach is then to use finite-order VAR inference or an appropriate bootstrap technique. Indeed, while \cite{inoueBootstrappingSmoothFunctions2002} have shown theoretically that the bootstrap is valid for sieve inference as $p$ grows with $T$, bootstrap resampling will simply approximate the empirical sample distribution of {finite-order} VAR($p$) IRFs.

\subsection{Practical Implications}

In applications the researcher is not able to control $T$, as it is a feature of the data, but they are able to calibrate $p$ when setting a VAR($p$) model and to choose $H$ when making inference on impulse responses. Accordingly, the important point to make is that sieve asymptotic results based on the assumptions of \cite{lewisPredictionMultivariateTime1985} might not show satisfactory inference properties in realistic economic settings which might be of interest. With monthly macroeconomic data, e.g. the well-known FRED-MD dataset \citep{mccrackenFREDMDMonthlyDatabase2015}, rules-of-thumb often suggest to set $p$ high enough to encompass at least a year of lags ($p \geq 12$) in order to capture simple calendar features of the data.\footnote{
	In practice, lag selection criteria like the Akaike (AIC) or the Bayesian (BIC) information criteria are often used to select $p$ in a data-driven way. For a thorough discussion of VAR model selection see for example Chapter 2, \cite{KilianLutkepohl2017}.
} In such applications, impulse responses to (structural) shocks can have economically relevant dynamics for horizons up to at least $H \geq 36$. Yet, to perform correct \textit{sieve} VAR inference using valid large-sample arguments, one would require $p = H$ and thus roughly $T \propto H^3$ at a minimum. This means that in relatively small samples sieve inference at long horizons could be problematic. Further, when evaluating model specifications with different $p$'s a strong implicit assumption on its relationship to $T$ would also be made. The researcher should not be unaware of these concerns when considering the sieve framework. 

\section{Monte Carlo Evidence}

\begin{figure}[t!]
	\centering
	\begin{subfigure}[b]{0.9\textwidth}
		\centering
		\makebox[\textwidth][c]{%
			\includegraphics[width=\linewidth]{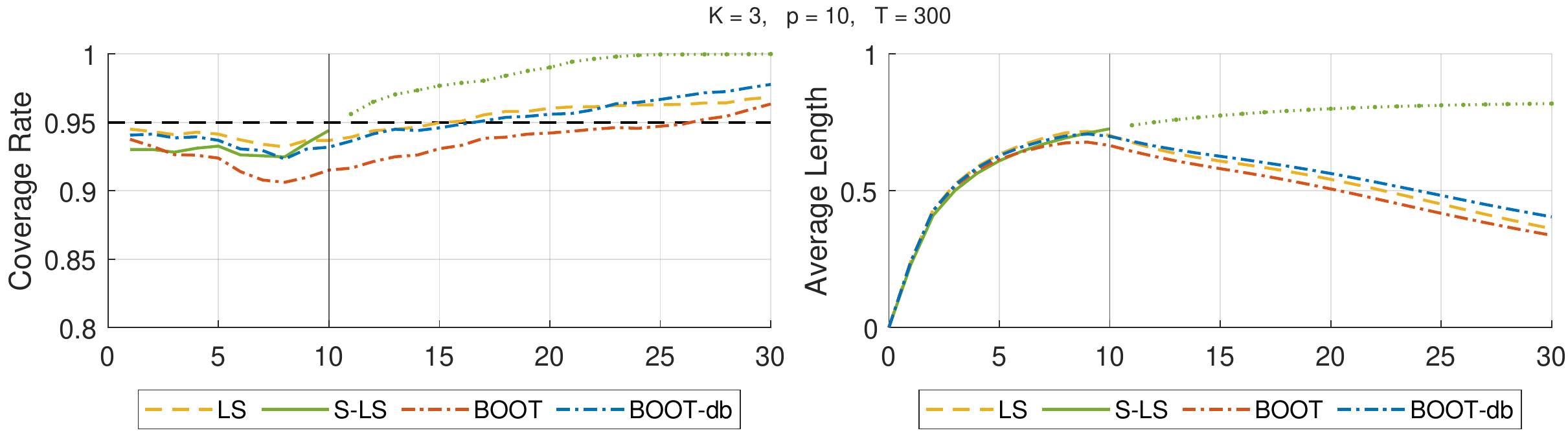}
		}
	\end{subfigure}\\
	\vspace{1em}
	\begin{subfigure}[b]{0.9\textwidth}
		\centering
		\makebox[\textwidth][c]{%
			\includegraphics[width=\linewidth]{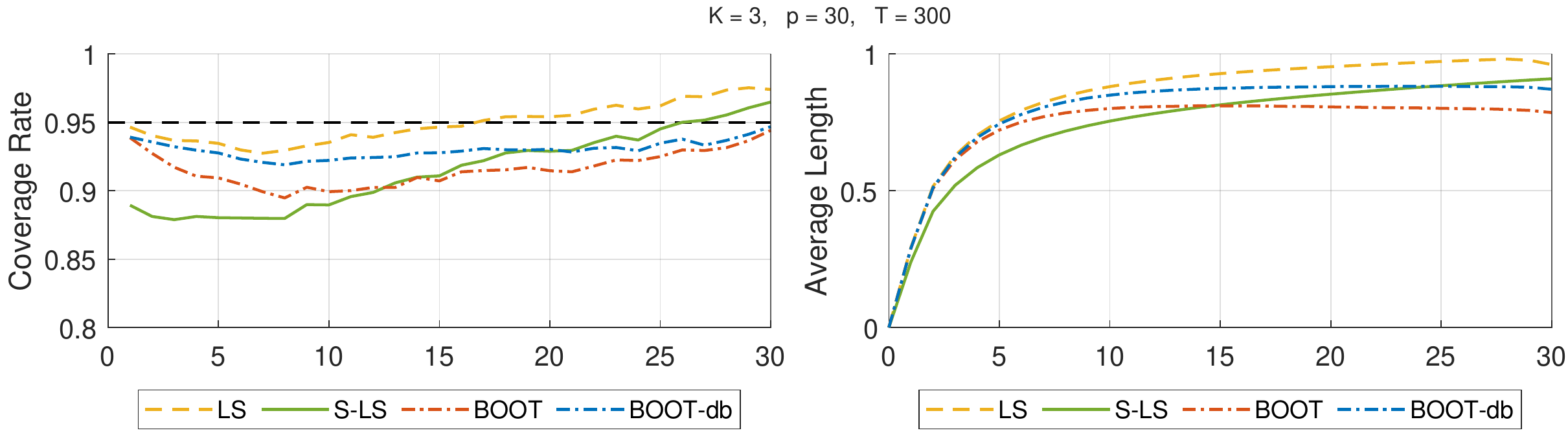}
		}
	\end{subfigure}\\
	\vspace{1em}
	\begin{subfigure}[b]{0.9\textwidth}
		\centering
		\makebox[\textwidth][c]{%
			\includegraphics[width=\linewidth]{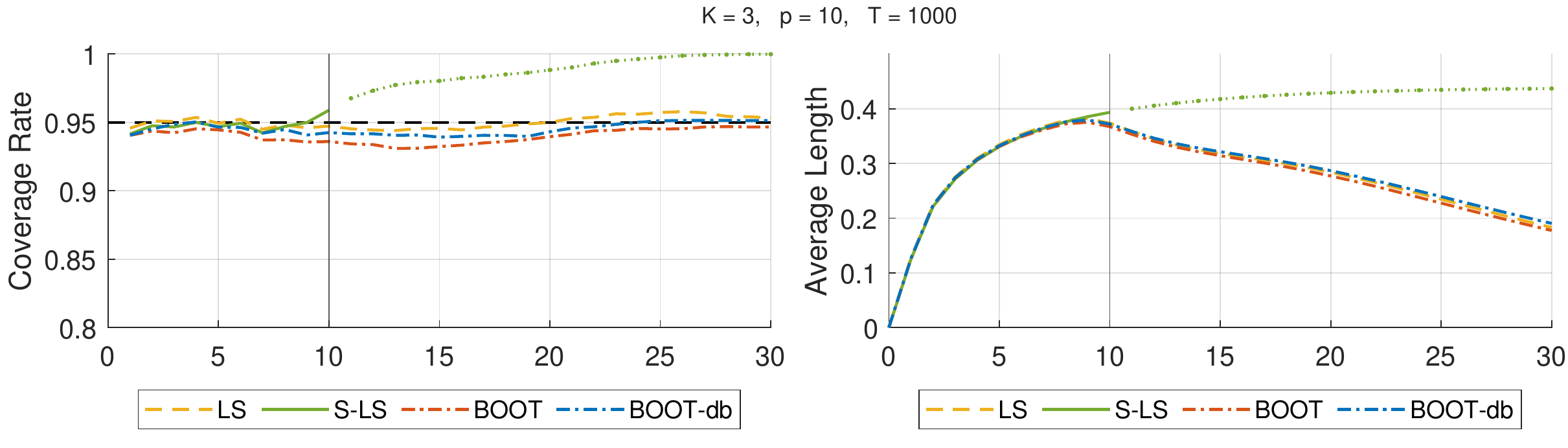}
		}
	\end{subfigure}\\
	\vspace{1em}
	\begin{subfigure}[b]{0.9\textwidth}
		\centering
		\makebox[\textwidth][c]{%
			\includegraphics[width=\linewidth]{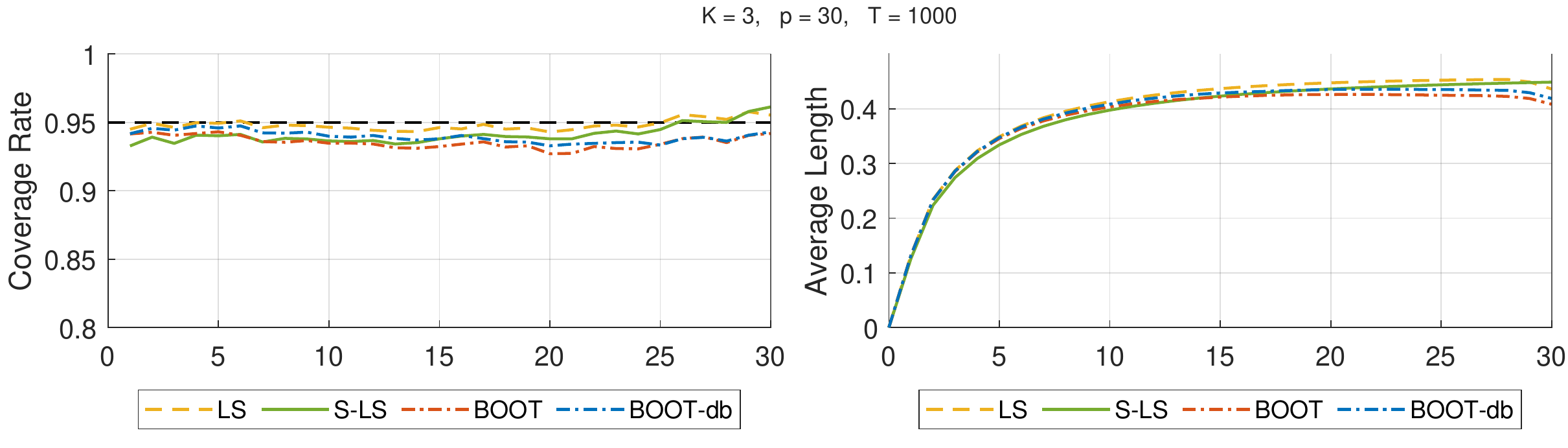}
		}
	\end{subfigure}\\
	\vspace{1em}
	\caption{Monte Carlo IRF confidence intervals simulation. Finite-order (LS), sieve (S-LS), bootstrap (BOOT) and de-biased boostrap-after-boostrap (BOOT-db) methods are shown. $p = \{10,\, 30\}$ lag lengths, $T = \{300,\, 1000\}$ sample sizes, $1000$ Monte Carlo replications. }
	\label{fig:mc_covg_length_10_30}
	%
\end{figure}

I use a simple Monte Carlo experiment -- inspired by the one employed by \cite{inoueBootstrappingSmoothFunctions2002} to produce Figure \ref{fig:inouekilian2002fig1} -- to showcase that sieve and finite-order VAR impulse response inference is asymptotically equivalent for all horizons $i \leq p-1$. This is also true for bootstrap inference produced by either the standard bootstrap or the de-biased boostrap-after-bootstrap proposed by \cite{kilianSmallsampleConfidenceIntervals1998}.\footnote{
	I use $M = 300$ boostrap replications, and for the de-biased bootstrap-after-boostrap I apply the "shortcut" to avoid nested loops, as suggested in \cite{kilianSmallsampleConfidenceIntervals1998}.
}

To simulate a VAR($\infty$) data-generating process, I construct samples from the VARMA(1,1) model specified in Appendix A.2, \cite{inoueBootstrappingSmoothFunctions2002}. Since for the purposes of this paper it is not necessary to study structural IRFs, I ignore the additional structure of the error matrix $\Sigma_u$ and instead draw $u_t$ from a standard multivariate normal distribution. Figure \ref{fig:mc_covg_length_10_30} plots coverage rates and lengths of (non-structural) impulse responses CIs constructed by fitting a VAR($p$) model via least squares. Finite-order, sieve, bootstrap and boostrap-after-boostrap confidence intervals are compared. The nominal confidence level is set to $95\%$, and both coverage and length are averaged across impulse responses. This is valid since the purpose is comparison and not inference. I choose two sample sizes, $T = 300$ and $T = 1000$, to better highlight the practical differences between methods, which are starker in small samples. A vertical solid line is added to indicate the threshold $i = p$.

Figure \ref{fig:mc_covg_length_10_30} clearly shows that sieve and finite-order confidence intervals both agree on the CI length for $i < p$ in large samples, although they differ meaningfully in small samples. Differences become more noticeable whenever $p$ is large since the sieve asymptotic approximation for the variance of $\Phi_i$ becomes less accurate, primarily because $T$ is not large enough compared to $p$. In fact, when $p = H = 30$ and $T = 300$, sieve CIs are thinner than finite-order CIs at all horizons, but wider than bootstrap CIs for large $i$. It seems, therefore, that no absolute ranking in terms of either coverage or length can be made between different CIs. In contrast, one can see that when $T = 1000$ there is little difference between methods even if $p = H$. Finally, since bootstrap and bootstrap-after-bootstrap methods are based on the (empirical) sample distribution of VAR($p$) IRFs, their length follows closely that of finite-order VAR CIs. This broadly agrees with the coverage results of Figure \ref{fig:inouekilian2002fig1} from \cite{inoueBootstrappingSmoothFunctions2002}, and its features can now be immediately explained.

\begin{figure}[t!]
	\centering
	\begin{subfigure}[b]{0.9\textwidth}
		\centering
		\makebox[\textwidth][c]{%
			\includegraphics[width=\linewidth]{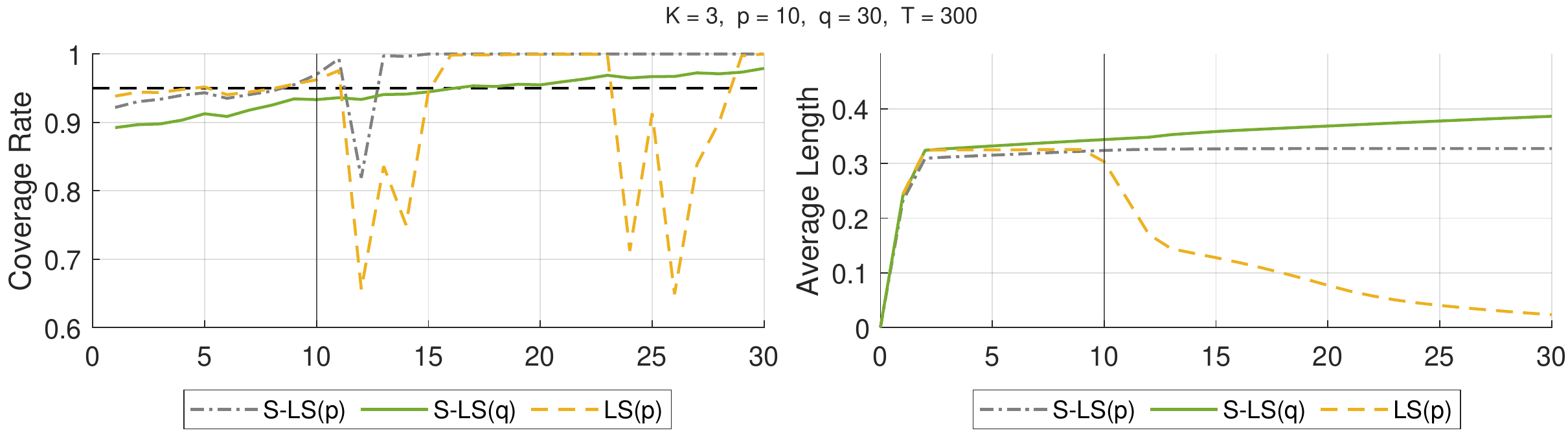}
		}
	\end{subfigure}\\
	\vspace{1em}
	\caption{Monte Carlo IRF simulation as counterexample for ill-constructed sieve CIs. Finite-order (LS) VAR($10$), sieve (S-LS) VAR($10$), and sieve VAR($30$) methods are shown. $T = 300$ sample size, $1000$ Monte Carlo replications. }
	\label{fig:mc_counterex_10_30}
	%
\end{figure}

To conclude, I present a counterexample to the seemingly "good" results of both finite-order and sieve VARs inference in the extrapolation regime $i > p$ of Figure \ref{fig:mc_covg_length_10_30}. The counterexample is based on a straightforward modification of the setup from \cite{inoueBootstrappingSmoothFunctions2002}. Let $A_1$ be the AR matrix from the VARMA(1,1) DGP considered above, and let
\begin{equation*}
	A_1^* := (A_1,\, \underbrace{0, \ldots, 0}_{\text{10 times}}\, , A_1/5, 0, A_1/10)
\end{equation*}
be defined as the modified AR coefficient matrix, such that $A_1^*$ has its largest companion matrix eigenvalue very close to that of $A_1$.\footnote{
	It is easy to check numerically that $\max_{j}
	\abs{\lambda^c_j(A_1)} = 0.895$ and $\max_{j}
	\abs{\lambda^c_j(A_1^*)} = 0.909$, where $\{\lambda^c_j(A)\}$ are the eigenvalues of the companion-form matrix of $A$.
} As in the previous Monte Carlo experiment, a VAR($\infty$) process is simulated, but this time inverting a VARMA(14,1) where $A_1^*$ takes the place of $A_1$. Figure \ref{fig:mc_counterex_10_30} compares the results of sieve and finite-order inference in this setup. When only $p = 10$ lags are estimated, the trailing non-zero coefficients in $A_1^*$ are ignored and the resulting inference is incorrect. Most importantly, it becomes clear that the non-diminishing length of sieve VAR(10) confidence intervals does not "insure" against lag under-estimation in any meaningful (or theoretically justifiable) way. The roots of this impropriety can be traced, in both sieve and finite-order asymptotic CIs, to being centered around the same point-wise impulse response estimates. Further, the unit coverage behavior of VAR(10) models around $i = 20$ of Figure \ref{fig:mc_covg_length_10_30} proves that there exist situations in which ill-constructed sieve confidence intervals \textit{may simultaneously yield {under-} and {over-coverage} of the true IRFs} at extrapolation horizons. Thus, in practice, sieve inference should not be seen as a panacea to finite-order model misspecification, but rather as an alternative asymptotic framework with additional limit assumptions on the model and its least-squares estimator.

\newpage
	
{\small\bibliography{./s_sample_sieve_arxiv}}
	
	
	
	
	
\end{document}